\documentclass[%
 notitlepage,
superscriptaddress,
reprint,
showpacs,preprintnumbers,
nofootinbib,
 amsmath,amssymb, 
 aps,
 prd,
 longbibliography,
]{revtex4-1}

\usepackage{cancel}
\usepackage{accents}
\usepackage{mciteplus,slashed}
\usepackage{amssymb,cancel,amsmath,relsize}
\usepackage{mathrsfs} 
\usepackage{dcolumn}
\usepackage{bm}
\usepackage[caption=false]{subfig}
\usepackage{appendix}
\usepackage{physics}
\usepackage{feynmp-auto}
\usepackage[T1]{fontenc}	
\usepackage{csvsimple}
\usepackage{hyperref}
\usepackage[section]{placeins}
\usepackage[capitalise]{cleveref}
\usepackage{booktabs}
\usepackage{graphicx}
\usepackage{mathrsfs}
\graphicspath{{Figures/}}

\usepackage[dvipsnames]{xcolor}
\usepackage[normalem]{ulem}

\usepackage{fontawesome} 

\def\iu{\mathrm{i}}
\def\e{\mathrm{e}}

\def\Jcc{\hat{J}_\nu}
\def\Jem{\hat{\mathscr{J}_\mu}}
\def\Javg{ \langle \hat{\mathscr{J}}_\mu \rangle}
\def\AmuAtom{A\mu,\text{atom}}
\def\MoneS{M_{\text{atom}}}

\newcommand{\sumsq}[1]{\left\langle |{#1}|^2\right\rangle} 

\DeclareMathOperator*{\sumint}{%
\mathchoice%
  {\ooalign{$\displaystyle\sum$\cr\hidewidth$\displaystyle\int$\hidewidth\cr}}
  {\ooalign{\raisebox{.14\height}{\scalebox{.7}{$\textstyle\sum$}}\cr\hidewidth$\textstyle\int$\hidewidth\cr}}
  {\ooalign{\raisebox{.2\height}{\scalebox{.6}{$\scriptstyle\sum$}}\cr$\scriptstyle\int$\cr}}
  {\ooalign{\raisebox{.2\height}{\scalebox{.6}{$\scriptstyle\sum$}}\cr$\scriptstyle\int$\cr}}
}

\begin{document}

\title{The high energy spectrum of internal positrons from radiative muon capture on nuclei}

\author{Ryan Plestid}
\email{rpl225@uky.edu}
\affiliation{Department of Physics and Astronomy, University of Kentucky,  Lexington, KY 40506, USA}
\affiliation{Theoretical Physics Department, Fermilab, Batavia, IL 60510,USA}
\author{Richard J. Hill}
\email{richard.hill@uky.edu}
\affiliation{Department of Physics and Astronomy, University of Kentucky,  Lexington, KY 40506, USA}
\affiliation{Theoretical Physics Department, Fermilab, Batavia, IL 60510,USA}

\date{\today}

\preprint{FERMILAB-PUB-20-525-T}

\begin{abstract}
{\centering{\href{https://github.com/ryanplestid/RMC-int-ext}{\large\color{BlueViolet}\faGithub}}  \\}  
    The Mu2e and COMET collaborations will search for  nucleus-catalyzed muon conversion to positrons ($\mu^-\rightarrow e^+$) as a signal of lepton number violation. A key background for this search is radiative muon capture where either: 1) a real photon converts to an $e^+ e^-$ pair ``externally" in surrounding material; or 2) a virtual photon mediates the production of an $e^+e^-$ pair ``internally''. If the $e^+$ has an energy approaching the signal region then it can serve as an irreducible background. In this work we describe how the near end-point internal positron spectrum can be related to the real photon spectrum from the same nucleus, which encodes all non-trivial nuclear physics.
\end{abstract}

 \maketitle 
 
 \section{Introduction}

 Charged lepton flavor violation (CLFV) is a smoking gun signature of physics beyond the Standard Model (SM)  and is one of the most sought-after signals  at the intensity frontier~\cite{Kuno:1999jp,Marciano:2008zz,Bernstein:2013hba,deGouvea:2013zba}. Important search channels involving the lightest two lepton generations are $\mu \rightarrow 3e$, $\mu\rightarrow e\gamma$, and nucleus-catalyzed $\mu\rightarrow e$ \cite{Kuno:1999jp,Bernstein:2013hba,deGouvea:2013zba,Marciano:2008zz,Calibbi:2017uvl,Galli:2019xop}. The current best limits on $\mu\rightarrow e$ come from SINDRUM-II \cite{Kaulard:1998rb,Bertl:2006up}.
 
 The upcoming Mu2e~\cite{Bernstein:2019fyh} and COMET~\cite{Lee:2018wcx} experiments will either probe or discover CLFV at unprecedented levels of precision. Both experiments expect on the order of $10^{18}$ muon capture events, and plan to measure the ratio, $R_{e\mu}$, of $\mu\rightarrow e$ events to total muon captures, at the level of $10^{-17}$.  

 \begin{figure}\centering
    \subfloat[][]{\includegraphics[height=0.33\linewidth]{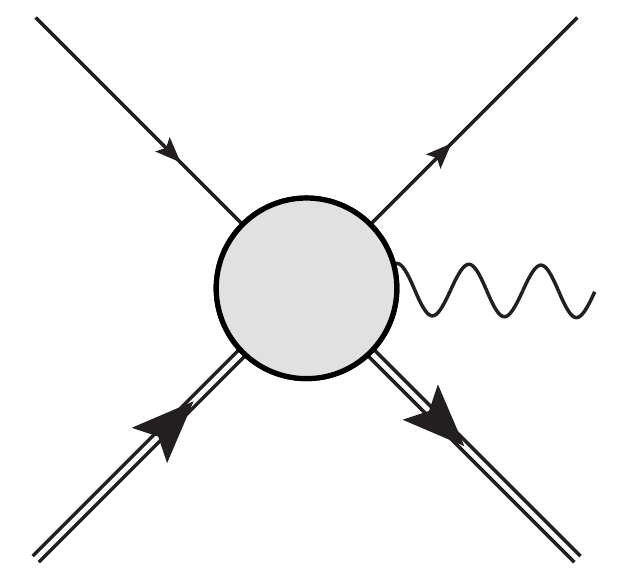}}%
    \qquad
    \subfloat[][]{\hspace{20pt} \includegraphics[height=0.33\linewidth]{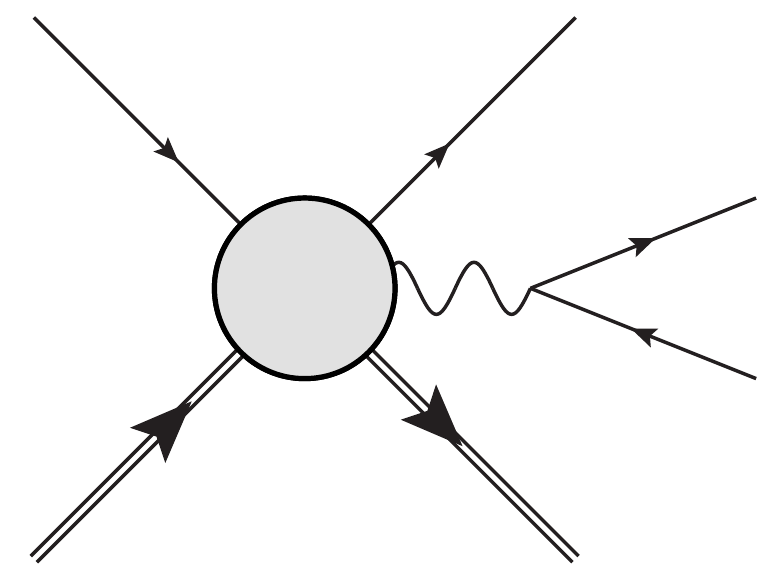}}
    \caption{Radiative muon capture on a nucleus (double lines) resulting in (a) a real photon and (b) a virtual photon that mediates the production of an electron positron pair. In this work we study how the $e^+$ (or $e^-$) spectrum in (b) is related to the photon spectrum in (a). \label{int-vs-ext}}
 \end{figure}

 Importantly, the Mu2e setup is ``charge symmetric'' such that the detection efficiencies for electrons and positrons are comparable. Thus, Mu2e will serve as a testing ground not just for the discovery of CLFV, but also for the discovery of lepton \emph{number} violation (LNV). Specifically, on a nucleus (e.g.\ aluminum), the reaction 
 \begin{equation}
    \mu^-  + [A,Z] \rightarrow e^+ + [A,Z-2]~,
 \end{equation} 
 becomes a viable target for observation (see also \cite{Kaulard:1998rb}). 
 
 While neutrinoless double beta ($0\nu\beta\beta$) decay is often touted as the most promising direction for the discovery of LNV, there do exist extensions of the SM that predict a more competitive signal in the $\mu e$ sector than in the $ee$ sector (cf. \cite{Geib:2016atx,deGouvea:2019xzm,Berryman:2016slh} and references therein). There also remains the looming possibility that $m_{ee}\approx 0$ (cf. Fig.~4 of \cite{Avignone:2007fu}), rendering $0\nu\beta\beta$ insensitive to LNV if it is mediated by a light Majorana neutrino. The fact that Mu2e is charge symmetric by design yields a new handle on LNV ``for free''. 
 
 Unfortunately, the charge-changing nature of $\mu^-\rightarrow e^+$ can result in a substantially lower positron energy compared to the electron energy in the CLFV channel $\mu^-\rightarrow e^-$. This is driven by the mass difference between $[A,Z]$ and $[A,Z-2]$~\cite{Yeo:2017fej}, and causes the signal region to approach, or overlap with, poorly-understood SM backgrounds.  The most important of these backgrounds is radiative muon capture (RMC),
 \begin{equation}\label{reaction}
    \mu(k) + [A,Z](p) \rightarrow \nu_\mu(k') + [A,Z-1](p') + \gamma(q) \,,
 \end{equation}
 where, for definiteness, we assume the final state nucleus is in its ground state\footnote{This assumption can be easily relaxed and a generic set of final states summed over as discussed in \cref{inclusive-X}.} (in the case of $^{27}$Al the final state would be $^{27}$Mg). 
 
 In searches for CLFV or LNV, incoming muons are the  source of both the $\mu \to e$ signal and the RMC background.  It is therefore critical to constrain  the energy spectrum of electrons and positrons from RMC in order to discriminate signal from background.

 There are two ways that RMC can contaminate the signal window in a CLFV or LNV search: 
 \begin{enumerate}
    \item External conversion: a (real) photon is produced and interacts with surrounding material ultimately pair-producing an electron positron pair.
    \item Internal conversion: a virtual photon mediates the production of an electron positron pair. 
 \end{enumerate}
 These two possibilities are shown schematically in \cref{int-vs-ext}. Both cases are subject to nuclear model dependence.\footnote{The RMC rate and spectral shape for capture on hydrogen has been studied in heavy baryon chiral perturbation theory~\cite{Meissner:1997mq,Ando:1997es}. 
Nuclear corrections substantially alter the spectral shape of RMC photons~\cite{Fearing:1988mu,Fearing:1991je,Bergbusch:1999ms,Cheoun:2003ah},
especially near the end-point \cite{Christillin:1979fa}.
}

 In the case of external conversion, nuclear model uncertainties can be circumvented at either Mu2e or COMET by directly measuring the real photon spectrum from RMC on aluminium. With this information in hand, dedicated Monte Carlo simulations (including the full detector geometry) can be used to predict the resultant electron and positron spectra. In this sense, the collaborations (Mu2e and COMET) control their own fate and can empirically constrain the external RMC backgrounds relevant for their own experiment. 
 
 Internal conversion is a more formidable challenge.  Early work by Kroll and Wada~\cite{Kroll:1955zu} investigated the ratio of $e^+e^-$ production relative to single photon production for a $1\rightarrow 2$ process. The results of their investigation showed, unsurprisingly, that the ratio is \emph{not} calculable without microscopic theoretical input. Rather, somewhat heuristically, they suggested that the infrared enhancement of the virtual photon favours small virtualities and argued for an approximation in terms of real photon matrix elements. 
 \begin{figure}\centering
    \includegraphics[width=\linewidth]{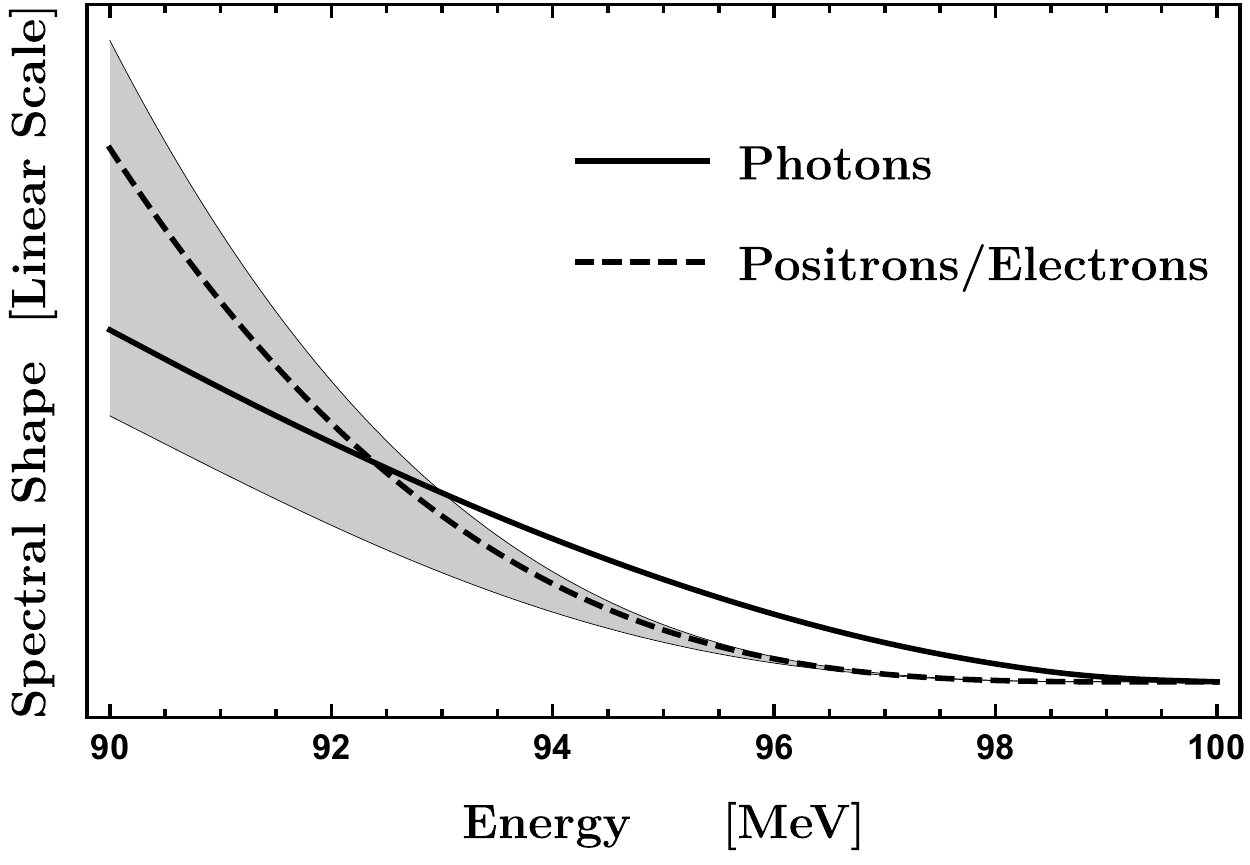}
    \caption{Spectrum of positrons/electrons as compared to real photons near the end-point assuming that the photon spectrum is given by $\dd \Gamma/ \dd E_\gamma \propto (1-2x+2x^2)x(1-x)^2$ with $x=E_\gamma/q_\text{max}$ and $q_\text{max}=100$ MeV. This corresponds to the closure approximation spectrum used in the analysis of RMC data on various nuclear isotopes in \cite{Bergbusch:1999ms}. The electron/positron spectrum is computed using \cref{P-int-dE,main-result} below. The spectrum is softened relative to photons because 1) the end-point is shifted by $m_e=0.511$ MeV, and 2) the probability of producing an $e^+e^-$ pair rises sharply as one moves further from the end-point as shown in \cref{main-plot}. Errors due to virtual-photon nuclear matrix elements (gray band) are estimated by treating $C_L$ and $C_T$ (see \cref{Error-Estimates}) as independent random variables drawn from a Gaussian distribution with unit variance. 
    \label{photon-vs-positron}    }
 \end{figure}

 In this work we critically re-examine this problem focusing specifically on the viability of using measurements of the (real) photon spectrum to predict the internal positron (or electron) spectrum. The main conclusion of our work is that the real photon spectrum is sufficient to predict the internal spectrum of positrons \emph{near the end-point} as shown in \cref{photon-vs-positron}.  This observation implies that if the real photon spectrum is measured, then the full (internal + external) electron and positron spectra can be predicted. If the photon spectrum cannot be measured, then Mu2e and COMET could use direct measurements of the total electron or positron spectrum to infer the RMC photon spectrum.

 The rest of the paper is organized as follows. In \cref{Nuclear-RMC} we define the  non-perturbative matrix element that governs the emission of real photons and discuss how this same matrix element also governs the emission of electron positron pairs at leading order in $\alpha$.  Next, in \cref{Real-Photon} we provide an explicit formula for the real photon spectrum, which naturally leads into \cref{Positron-Spectrum} where we provide the corresponding expression for the positron spectrum. In \cref{Near-End-Point} we discuss the near end-point spectrum and demonstrate how the real photon spectrum can be used to predict internally converted positrons. Finally, in \cref{Conclusions} we summarize our conclusions, and suggest future improvements.

 We also provide four appendices. In \cref{Error-Estimates} we discuss how one can parameterize sub-leading corrections to the near end-point approximation discussed above. In \cref{OFPT} we give a formal argument for some power counting details needed to justify the approximation of transversely polarized virtual photon matrix elements by their real photon counter parts. To facilitate a comparison between our work and that of Kroll and Wada we provide a short discussion of the correspondence in \cref{KW-vs-HP}. We also discuss our disagreements with their conclusions there.   Finally in \cref{inclusive-X} we describe how to generalize the analysis presented in the main text to the case of inclusive final states.

 \section{Radiative muon capture on a nucleus \label{Nuclear-RMC}} 
 
The relevant S-matrix element for radiative muon capture is the overlap of an in-state, $\ket{\AmuAtom }_\text{in}$  containing a muonic atom, with an out state $\ket{A'\nu_\mu \gamma}_\text{out}$ containing a recoiling nuclear system $A'$, a muon neutrino $\nu_\mu$, and a photon $\gamma$:  
\begin{equation}
    S_\gamma=  ~~_\text{out}\hspace{-3.5pt}\braket{A'\nu_\mu\gamma}{\AmuAtom}_\text{in}~.
\end{equation}
This can be expressed via the LSZ reduction formula as
\begin{equation}\begin{split}\label{Sgamma}
    S_\gamma&= (2\pi)^4\delta^{(4)}(\Sigma_3 P) \epsilon^\mu ~~ _\text{out}\hspace{-3.5pt}\mel{A'\nu_\mu}{\hat{\mathscr{J}}_\mu}{\AmuAtom}_\text{in}~,
\end{split}\end{equation}
where ${\hat{\mathscr{J}}_\mu}$ denotes the electromagnetic current and the notation $\Sigma_3 P$ signifies that this momentum conservation is for three on-shell particles in the final state.

At leading order in $\alpha$, the $S$ matrix for internal pair production is given by (we use  $\Javg \equiv$  $_\text{out}\hspace{-3.5pt}\mel{A'\nu_\mu}{\hat{\mathscr{J}}_\mu}{\AmuAtom}_\text{in}$ from here on for brevity's sake)
\begin{equation}\begin{split}\label{See}
    S_{ee}
    &=-(2\pi)^4\delta^{(4)}(\Sigma_4 P)  {\Javg}_*~ \frac{\sqrt{4\pi\alpha}}{m_*^2} \bar{u} \gamma^\mu v~.
\end{split}\end{equation}
Note that the in-out matrix element in (\ref{See}) is evaluated ``off-shell'', in that $m_*^2\equiv(p_++p_-)^2\neq 0$, where $p_\pm$ refer to the four momenta of $e^+$ and $e^-$ respectively. 

Stripping off the four-momentum conserving delta function, we can therefore identify the matrix elements as 
\begin{align}
    \label{eq6}
    \iu\mathcal{M}_\gamma &= {\Javg}_0  \epsilon^\mu~,\\
    \iu\mathcal{M}_{ee} &= -{\Javg}_*\frac{\sqrt{4\pi\alpha}}{m_*^2} \bar{u} \gamma^\mu v ,\label{eq7}
\end{align} 
where in \cref{eq7} we work at leading order in $\alpha$. The subscript reminds us whether the matrix element has been evaluated for real, ${\Javg}_0$, or virtual, ${\Javg}_*$, photon kinematics. 

To calculate the rate of decay we can make use of the standard formula\footnote{The conventional muon capture formula (see e.g.\ \cite{Fearing:1979be})  written in terms of $\abs{\psi_{1\text{s}}(0)}^2$ can be recovered in the non-relativistic limit by constructing the bound state $\ket{1\text{s},A\mu}$ out of plane-wave states $\ket{\mu(k)}$ and $\ket{A(-k)}$ as described in \S5 of \cite{Peskin:1995ev}.}  
\begin{equation}
    \Gamma = \frac{1}{2 \MoneS }  \int \dd \Phi_n \sumsq{\mathcal{M}}~,
\end{equation}
where $\MoneS$ is the mass of the muonic atomic (including binding energy), $\sumsq{\cdot}$ 
denotes an average over initial-state spins and a sum over final-state spins,
and $\dd \Phi_n$ is $n$-body Lorentz invariant phase space,
\begin{equation}
    \dd \Phi_n = \qty[ \prod_{i=1}^n \frac{\dd^3 p_i}{2E_i(2\pi)^3}  ]  \times  (2\pi)^{4}\delta^{(4)}(\Sigma_n P)~.
\end{equation}    
In what follows we consider the spin-averaged matrix element which is rotationally invariant. This allows us to choose our coordinate system to lie with the $\hat{z}$ axis along the direction of the photon's momentum without loss of generality. 
\section{Real photon spectrum \label{Real-Photon}}
Let us introduce the tensor
\begin{equation}
\mathscr{J}_*^{\mu\nu} = \sum_{\text{spins}} \langle \hat{\mathscr{J}}^\mu\rangle_* ~\times ~ \langle \hat{\mathscr{J}}^{\nu\dagger}\rangle_*~,
\end{equation} 
where an average over initial spin states, and sum over final spin states is performed.

The real photon matrix element squared, $\sumsq{\mathcal{M}_\gamma}$, can be expressed in terms of this tensor  as  
\begin{equation}
   \sumsq{\mathcal{M}_\gamma}=  (-g_{\mu\nu}^\perp) \mathscr{J}^{\mu\nu}_0~,
\end{equation}
where we have used $\sum \epsilon_\mu \epsilon_\nu^* = -g_{\mu\nu}^\perp$ for the sum over physical photon polarizations. The differential rate (or spectrum) of real photons in the lab frame is then given by 
\begin{equation}
    \dv{\Gamma_\gamma}{E_\gamma} = 
     \frac{1}{2 \MoneS }  \int \dd \Phi_3 \qty[ \mathscr{J}^{11}_0+ \mathscr{J}^{22}_0] \delta(q_0- E_\gamma)~.
\end{equation}

\section{Positron spectrum \label{Positron-Spectrum}} 

The matrix element for $e^+e^-$ creation can be expressed in terms of $\mathscr{J}_*^{\mu\nu}$ as 
\begin{equation}
    \sumsq{\mathcal{M}_{ee}} =  \frac{4\pi\alpha}{m_*^4}  L_{\mu\nu} \mathscr{J}_*^{\mu\nu}  ~, 
\end{equation}
where the lepton tensor $L_{\mu\nu}$ is defined as 
\begin{equation}
L_{\mu\nu} =  \mathrm{Tr}\qty[ (\slashed{p}_-+m_e)\gamma_\mu (\slashed{p}_+-m_e)\gamma_\nu]~. 
\end{equation}
The positron spectrum is then given by 
\begin{equation}\label{Gamma-ee}
    \dv{\Gamma_{ee}}{E_+} = 
     \frac{1}{2 \MoneS }  \int \dd \Phi_4  \frac{4\pi\alpha}{m_*^4}  L_{\mu\nu} \mathscr{J}_*^{\mu\nu}  \delta(\mathfrak{E}_+- E_+)~,
\end{equation}
where $\mathfrak{E}_+$ is a function of the phase space variables that will be specified explicitly below in \cref{Funny-E}. 

At this point, to make contact with the real photon spectrum, it is important to decompose phase space appropriately. This can be done via \cite{HitoshiPS}
\begin{equation}\label{phase-space-decomp-impl}
    \dd \Phi_4 = \dd \Phi_{3*} \frac{\dd m_*^2}{2\pi} \dd \Phi_2(\gamma_*\rightarrow e^+ e^-)~,
\end{equation}
where $\Phi_{3*}$ is the three body phase space for $A'$, $\nu_\mu$ and a particle of mass $m_*$. Let us compare to the massless case. We will explicitly integrate over the $A'$ system's momentum such that only the energy conserving delta function,  $\delta(\Sigma_{3*} E)$  and the neutrino-photon phase space remain 
\begin{align}
    \dd \Phi_{3*} &= \frac{1}{2 E_{A'}}
   \frac{\dd^3 k'}{2 k'  (2\pi)^3} \frac{\dd^3 q}{2 q_0  (2\pi)^3} (2\pi)\delta(\Sigma_{3*}E)
   \label{3star-phase-space}\\
   &= \frac{1}{2 E_{A'}}
   \frac{\dd^3 k'}{2 k'  (2\pi)^3} \frac{ \dd \Omega_q }{ (2\pi)^3} \times 
   \frac{\beta_* q_0}{2} \dd q_0 (2\pi)\delta(\Sigma_{3*} E)\nonumber~,
\end{align}
where $\beta_*=q/q_0=\sqrt{1-m_*^2/q_0^2}$, and we have made use of $\dd q/\dd q_0 = 1/\beta_*$. This is to be compared with the case of a massless photon which does not have the factor of $\beta_*$
\begin{equation}\label{3-phase-space}
\dd \Phi_{3}=\frac{1}{2 E_{A'}}
   \frac{\dd^3 k'}{2 k'  (2\pi)^3} \frac{ \dd \Omega_q }{ (2\pi)^3} \times 
  \frac{q_0}{2} \dd q_0 (2\pi)\delta(\Sigma_3 E)~.
\end{equation}
The energy conservation condition is (for a recoiling $A'$ of mass $M$)
\begin{equation}\label{energy-conservation}
  \sum_{3*} E=   k+q_0 + \underbrace{\sqrt{k^2 + |\vec{q}|^2 + 2|\vec{q}|k\cos\theta_{\nu\gamma}  + M^2}}_{E_{A'}}- \MoneS~,
\end{equation}
where $|\vec{q}|^2 = q_0^2-m_*^2$ and $\theta_{\nu\gamma}$ is the opening angle between the photon and neutrino in the lab frame. Integrating over the neutrino energy introduces a factor of $\qty[ \pdv{\Sigma_{3*} E}{k}]^{-1}$:
\begin{equation}
    \frac{1}{1+\frac{k + |\vec{q}|\cos\theta}{E_{A'} }} = \frac{E_A'}{\MoneS - (1-\beta_*\cos\theta_{\nu\gamma}) q_0} ~.
\end{equation}
 The factor of $E_{A'}$ cancels against the factor of $E_{A'}$ in the denominator of \cref{3-phase-space,3star-phase-space} such that the phase space for a massive photon can be related to the phase space for a massless photon via
\begin{equation}\label{3-vs-3*}
    \dd \Phi_{3*}= \dd \Phi_3 \times \beta_* F_*~,
\end{equation}
with $\Phi_{3}$ being independent of $m_*$ and $\dd \Phi_2$, and the variable $F_*$ being given by (cf. Eq.\ (2) of \cite{Fearing:1979be})
\begin{equation}\begin{split}
    F_* &=\qty[\qty(\pdv{\Sigma_{3} E}{k})\bigg/ \qty(\pdv{\Sigma_{3*} E}{k}) ] \times \frac{k'_*}{k'}\\
    &= \frac{\MoneS - (1-\cos\theta_{\nu\gamma}) q_0}{\MoneS - (1-\beta_*\cos\theta_{\nu\gamma}) q_0}~ \times \frac{k'_*}{k'} ~,
\end{split}\end{equation} 
where the various energetic factors in the phase space measure are modified so that they satisfy \cref{energy-conservation}.

In terms of these variables we can then write 
\begin{equation}\label{phase-space-decomp-expl}
   \int \dd \Phi_4 = \int \dd\Phi_3 \int_{4m_e^2}^{q_0^2} \frac{\dd m_*^2}{2\pi} \beta_* F_* \int \dd \Phi_2~,
\end{equation}
with $\Phi_2$ the two-body phase space for a virtual photon of mass $m_*$ decaying into an $e^+e^-$ pair. We have inserted the integral symbols explicitly to emphasize the order in which the they must be performed. 

Let us now study \cref{Gamma-ee}, performing the integrations sequentially from right to left as suggested by \cref{phase-space-decomp-expl}. Since the decomposed phase space is factorized into independently Lorentz invariant pieces we can carry out the integration in the frame of our choice. The quantity $\mathscr{J}_{\mu\nu}$ is independent of the two-body phase space $\Phi_2$ and so we can first evaluate 
\begin{equation}
    \int \dd \Phi_2 L_{\mu\nu} \delta(\mathfrak{E}_+ -E_+) ~,
\end{equation}
in the rest frame of the photon. Two body phase space in this frame is given by 
\begin{equation}
    \dd \Phi_2 = \frac{\dd \cos\vartheta \dd \varphi }{32\pi^2} \beta_e~ ,
\end{equation}
where $\beta_e=\sqrt{1-4 m_e^2/m_*^2}$ and coordinates are defined such that $\hat{z}$ is parallel to the boost direction that connects the photon's rest frame to the lab frame.  As already discussed above, this choice can be made without loss of generality because of the spherical symmetry of the spin-averaged matrix element $\sumsq{\mathcal{M}}$. 

In terms of these variables the function $\mathfrak{E}_+$ that determines the positron's energy is given by 
\begin{equation}\label{Funny-E}
    \mathfrak{E}_+(q_0, \cos\vartheta)= 
    \frac{q_0}{2} (1+ \beta_e \beta_*\cos\vartheta)~,
\end{equation}
where $\beta_*=\sqrt{1-m_*^2/q_0^2}$ is (as above) the velocity of the virtual photon.

The delta function is independent of $\varphi$ and so it is convenient to define $\langle L_{\mu\nu} \rangle_\varphi= \int \frac{\dd \varphi}{2\pi}  L_{\mu\nu}$ given explicitly by 

\begin{widetext}
\begin{equation}\label{lepton-tensor}
    \langle L_{\mu\nu} \rangle_\varphi= m_*^2  \begin{pmatrix}
         0 & 0 & 0 & 0 \\     
         0 & 2-\left(1-\cos^2\vartheta\right) \beta_e^2 & 0 & 0 \\ 
         0 & 0 & 2-\left(1-\cos^2\vartheta\right) \beta_e^2 & 0 \\ 
         0 & 0 & 0 & 2 \left(1-\cos^2\vartheta \beta_e^2\right)  \\       
    \end{pmatrix}
    \qq{(in photon rest frame)}~.
\end{equation}
\end{widetext}
Then we find 
\begin{equation}
   \int \dd \Phi_2 L_{\mu\nu} \delta(\mathfrak{E}_+ -E_+)=  \left. \frac{\beta_e}{8\pi} \frac{1}{q_0 \beta_e \beta_*} \langle L_{\mu\nu} \rangle_\varphi \right\rvert_{\cos\theta}~,
\end{equation}
where we have used (note the different fonts of $\vartheta$ vs $\theta$)
\begin{equation}
\delta(\mathfrak{E}_+ -E_+)= \frac{2}{q_0 \beta_* \beta_e} \delta(\cos\vartheta-\cos\theta)~.  
\end{equation} 
The emission angle of the positron, $\cos\theta$, is found by solving $E_+=\mathfrak{E}_+(q_0,\cos\theta)$,  being given explicitly by
\begin{equation}\label{cos-theta-func}
     \cos\theta(E_+,q_0,m_*)=\frac{ 2 E_+/q_0 -1}{\beta_*\beta_e} . 
\end{equation}
Obviously,  $\cos\theta$ must lie in the interval $[-1,1]$ and so Heaviside functions enforcing this condition should be included:
\vfill
\begin{widetext}
\begin{align}\label{exact-rate-1}
    \dv{\Gamma_{ee}}{E_+}&= \frac{1}{2 \MoneS} \int \dd \Phi_3  ~\frac{\dd m_*^2}{2\pi} \times \beta_* F_* \times \frac{4\pi \alpha}{m_*^4}  \times \frac{1}{8\pi}\frac{1}{q_0 \beta_*} \langle L_{\mu\nu}\rangle_\varphi \mathscr{J}^{\mu\nu}_*
    \Theta(1-\abs{\cos\theta})\nonumber\\
    &= \frac{1}{2 \MoneS} \int \dd \Phi_3 \frac{\alpha}{4 \pi q_0}\int \dd m_*^2~ F_*\qty[ \frac{1}{m_*^4} \langle L_{\mu\nu}\rangle_\varphi \mathscr{J}^{\mu\nu}_* ] \Theta(1-\abs{\cos\theta})\\
    &=\frac{1}{2 \MoneS} \int \dd \Phi_3 \frac{\alpha}{4 \pi q_0}  \int \dd m_*^2 \frac{F_*}{m_*^2}~\qty[\qty(2-\qty[1-\cos^2\theta] \beta_e^2)\qty(\mathscr{J}^{11}_* + \mathscr{J}^{22}_*) +2\left(1-\cos^2\theta \beta_e^2\right) \mathscr{J}^{33}_*]\Theta(1-\abs{\cos\theta}) \nonumber~. 
\end{align}
\end{widetext}
In \cref{exact-rate-1} we have contracted  $\mathscr{J}_*^{\mu\nu}$ against the lepton tensor evaluated in the rest frame of the photon.\footnote{If the longitudinal matrix element is evaluated in the lab frame then one finds  $(m_*/q_0) \times[L_*^2]_\text{lab}$ \cite{Kroll:1955zu}. We can relate $[L_*^2]_\text{lab}$ to $[L_*^2]_\text{rest}$ by boosting between frames, and one finds that $[L_*^2]_\text{lab}=(q_0/m_*)\times [L_*^2]_\text{rest}$ such that the factor of $m_*/q_0$ is precisely canceled.}

Now let us define auxiliary variables to condense the notation:
\begin{align}
    T_*^2 &= \frac{[\mathscr{J}^{11}_* + \mathscr{J}^{22}_*]- [\mathscr{J}^{11}_0 + \mathscr{J}^{22}_0]}{\mathscr{J}^{11}_0 + \mathscr{J}^{22}_0}~,\\
     L_*^2 &= \frac{\mathscr{J}^{33}_*}{\mathscr{J}^{11}_0 + \mathscr{J}^{22}_0}~.
\end{align}
The quantities $T_*^2$, and $L_*^2$ are related to the transverse and longitudinal matrix elements respectively. Notice that $T_*^2(m_*=0) =0$ by construction such that $T_*^2\sim O(m_*^2)$ in the small $m_*$ limit. 

Having introduced all of the necessary ingredients, we  can now write the positron spectrum as if it were produced by real photons converting ``internally'' via a probabilistic process described by a function $P_\text{int}$ 
\begin{equation}\begin{split}
\label{P-int-Phi3}
    \dv{\Gamma_{ee}}{E_+}&= \frac{1}{2 \MoneS} \int \dd \Phi_3 \qty[\mathscr{J}^{11}_0 + \mathscr{J}^{22}_0 ] P_\text{int}(E_+|q_0, \Pi)~.
\end{split}\end{equation}
The symbol $\Pi$ represents a set of 3-body phase space variables (e.g. the photon-neutrino opening angle $\cos\theta_{\nu\gamma}$), and we have introduced the internal conversion probability 
\pagebreak
\begin{widetext}
\begin{equation}\label{P-int-def}
     P_\text{int}(E_+|q_0, \Pi)=
    \frac{1}{q_0 } \frac{\alpha}{4\pi}\int \frac{\dd m_*^2}{m_*^2} F_* \qty[\qty(2-\qty[1-\cos^2\theta] \beta_e^2)(1+T_*^2) +2\left(1-\cos^2\theta \beta_e^2\right) L_*^2]\Theta(1-\abs{\cos\theta})~,
\end{equation}
\end{widetext}
where $ \dd q_0\times  \int P_\text{int}(E_+|q_0, \Pi)\dd\Pi$ can be interpreted as the probability for photons with energy between $q_0$ and  $q_0+\dd q_0$ to produce an electron positron pair.  

The functions $T_*^2$ and $L_*^2$ are unconstrained by measurements of the on-shell photon spectrum. As discussed above, $T_*^2\sim O(m_*^2)$ for small virtualities, and so if $m_* \ll q_0$ then we can expect it to be small. In contrast, $L_*$ is expected to be $O(1)$ in the $m_*\rightarrow0$ limit.\footnote{This is qualitatively different than the conclusion of \cite{Kroll:1955zu}. We discuss the origin of these differences in \cref{KW-vs-HP}.} Notice, however, that as $\cos\theta\rightarrow 1$ (or equivalently as $E_+\rightarrow q_0 -m_e$) that the longitudinal matrix elements are suppressed.  We will discuss this in more detail in \cref{Near-End-Point}. 

Our definition of $P_\text{int}$ still depends on the unconstrained matrix elements $L_*^2$ and $T_*^2$. As we will show in the next section there exists a limit where $L_*^2$ and $T_*^2$ can be neglected.  In this case \cref{P-int-Phi3} can be re-written by carrying out the integration over all of the 3-body phase space except for $\dd q_0 $ in which case we find
\begin{equation} \label{P-int-dE}
    \dv{\Gamma_{ee}}{E_+}= \int \dd E_\gamma \dv{\Gamma_\gamma}{E_\gamma} P_\text{int}(E_+|E_\gamma)~,
\end{equation}
which allows us to construct the positron spectrum using the measured photon spectrum from the same nucleus, and the calculable function $P_\text{int}(E_+|E_\gamma)$. We now turn our attention to the aforementioned limit in which $P_\text{int}$ is independent of $T_*^2$ and $L_*^2$.

\section{Near end-point spectrum \label{Near-End-Point}}

As discussed in the introduction, for $\mu\rightarrow e$ searches  it is the high-energy tail of the RMC spectrum that is most important. This motivates studying the internal conversion probability in the limit where $E_+\rightarrow E_+^{\text{(max)}}$. This is conveniently parameterized by the dimensionless ratio formed from the electron's kinetic energy to the virtual photon energy,
\begin{equation}
    \delta= \frac{q_0 - (E_+ + m_e)}{q_0}= \frac{E_--m_e}{q_0}~,
\end{equation}
which tends to zero as the positron approaches its endpoint.

\begin{figure}
    \includegraphics[width=\linewidth]{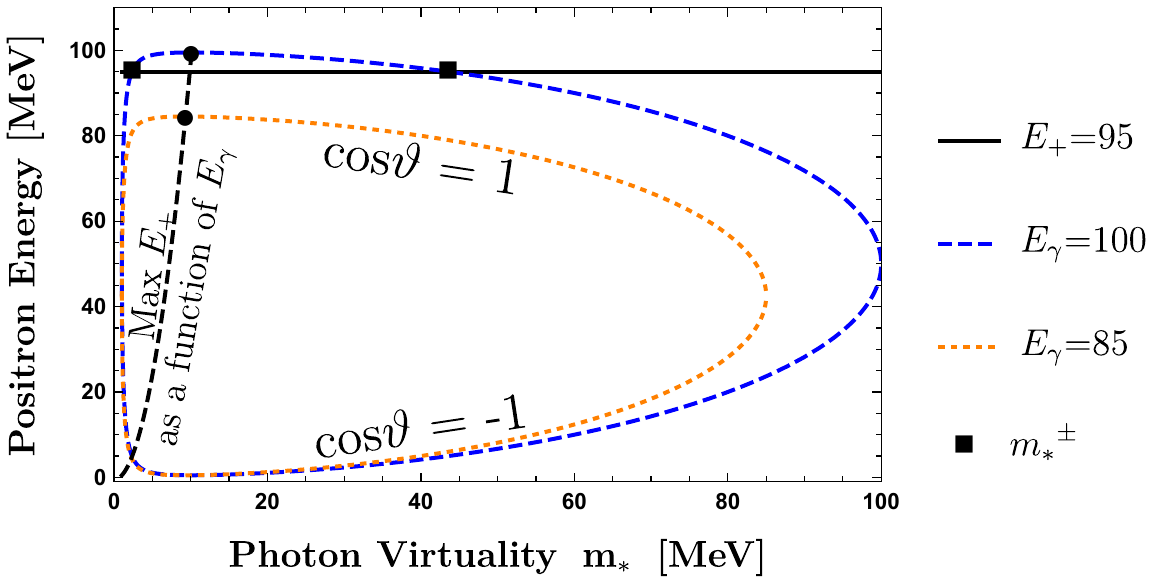}
    \caption{Phase space for radiative muon capture. Varying $\cos\vartheta$ from $-1$ to $1$ at fixed $m_*$ moves from the bottom of the hull to the top of the hull. Changing $E_\gamma= q_0$ changes the size of the hull to be integrated over (i.e. orange dotted curve [$E_\gamma=85$ MeV] $\rightarrow$ blue dashed curve [$E_\gamma=100$ MeV]). If $E_+\rightarrow q_0$ then only the top-tip of the hull contributes to the phase space integration. Units of $E_+$ and $m_*$ are in MeV. \label{phase-space-hull} }
\end{figure}

\begin{figure}\centering
    \includegraphics[width=\linewidth]{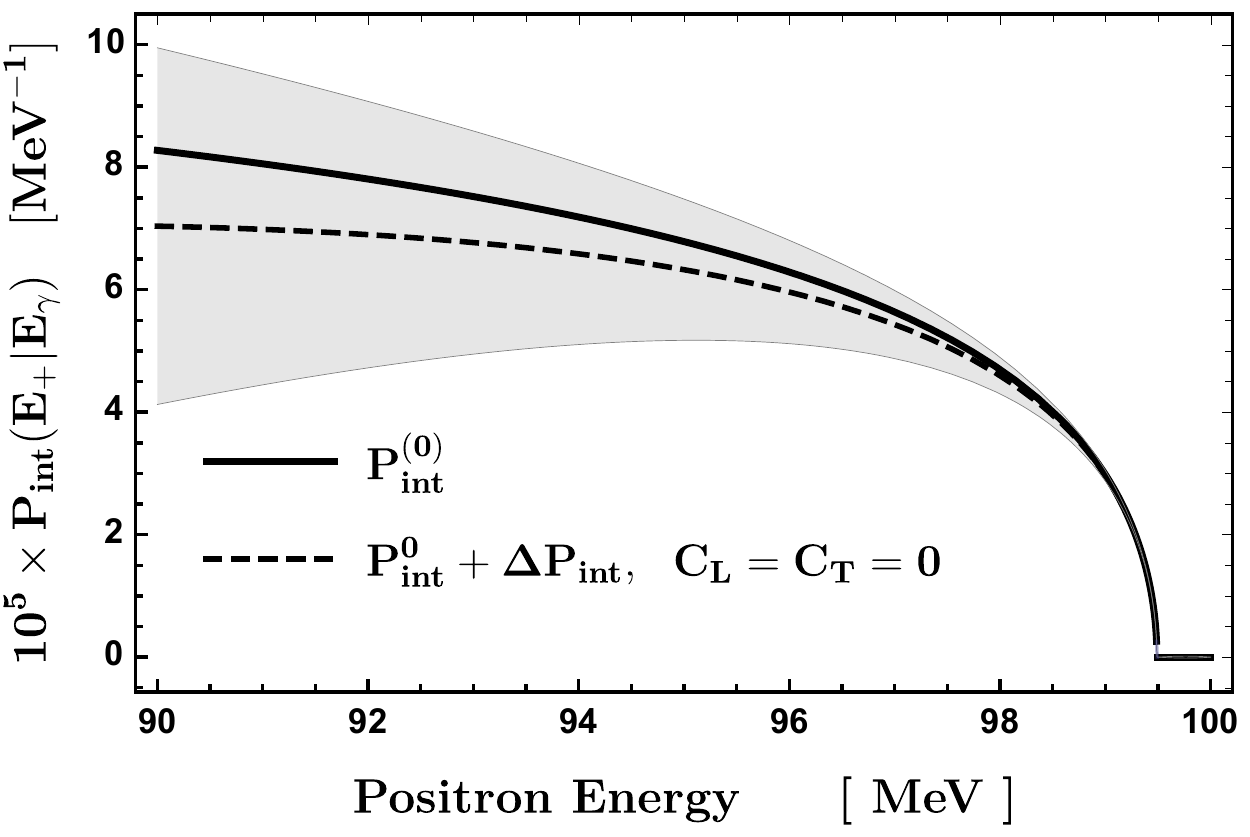}
    \caption{Probability per differential bin (i.e.\ of width $\dd E_\gamma$) of producing a positron with energy $E_+$ given a photon energy of $E_\gamma=100$ MeV, \cref{main-result,max-min-mstar}. We plot $P_\text{int}^{(0)}$ of \cref{p-int-approx} as well as error estimates (gray band) obtained by considering an ensemble of the two parameters $C_L$, and $C_T$, that capture nuclear structure dependence (see discussion in \cref{Error-Estimates}). We treat $C_L$ and $C_T$ as random variables drawn from independent Gaussian distributions such that $\langle C_L^2 \rangle= \langle C_T^2  \rangle =1$. \label{main-plot}}
 \end{figure}

Let us now study the integral over $m_*$ in \cref{P-int-def} more carefully. Having restricted ourselves to a specific value of $E_+$, the limits of integration on $m_*$ are supplied by the Heaviside function. This can be understood by plotting plotting $\mathfrak{E}_+$ for $\cos\theta=\pm 1$, as a function of $m_*$ at different fixed values of $q_0$ as depicted in \cref{phase-space-hull}.  At a fixed positron energy $E_+$, the limits of integration over $m_*$ can be found by solving
\begin{equation}
    \mathfrak{E}_+( m_*, q_0, \cos\vartheta=1)= E_+ ~.
\end{equation}
These are given explicitly by 
\begin{widetext}
\begin{equation}\begin{split}\label{max-min-mstar} 
     m_*^{(\pm)} =   \sqrt{2\qty[m_e^2+E_+( q_0-E_+) \pm \sqrt{\left(E_+^2-m_e^2\right) \left([q_0-E_+]^2-m_e^2\right)}]}    ~.  
\end{split}\end{equation}
\end{widetext}
Notice that $m_*^{(-)} \geq 2 m_e$ and consequently $m_e/m_* \leq 1$ over the full range of integration. If we introduce the small parameter 
\begin{equation}
    \epsilon=\frac{ m_e}{q_0}~,
\end{equation}
then the maximal value of $m_*$  is set by
\begin{equation}
    m_*^{(+)} = q_0 \qty[2\sqrt{(1-\delta ) \delta } +O(\epsilon)]~.
\end{equation}
We therefore find that $m_*/q_0 \leq O(\sqrt{\delta})$ in the small $\delta$ limit (provided $\epsilon \ll \delta$). Similarly, for a high-energy positron to emerge from a virtual photon, it must be emitted collinear to the photon's momentum, $\cos\theta\approx 1$.  Explicitly, 
\begin{equation}\begin{split}
  1- \cos\theta &=1-   \frac{ 2 E_+/q_0 -1}{\beta_*\beta_e}= O(\delta)~,
\end{split}\end{equation}
where we treat $\beta_e\approx \beta_*\approx 1$ (see \cref{Error-Estimates} for details). 
Thus, we can re-express the internal conversion probability for $E_+ \rightarrow q_0$ as 
\begin{equation}\begin{split}\label{p-int-approx}
     P_\text{int}&=
    \frac{1}{q_0 } \frac{\alpha}{\pi}\int_{m_*^{(-)}}^{m_*^{(+)}} \frac{\dd m_*^2}{m_*^2} F_* \\
    &\quad\quad\quad\quad\quad\quad\times \qty[2 +2 T_*^2- O(\delta)  +O(\delta)\times L_*^2]\\
    &= P_\text{int}^{(0)}(E_+|q_0) + O(\delta)~.
\end{split}\end{equation}
where we have counted $T_*^2\sim O(\delta)$ (see \cref{OFPT} for more details). The longitudinal matrix elements, by contrast, are taken to be $O(1)$ in the $m_*\rightarrow 0$ limit, but are suppressed by  $\left(\cos^2\theta \beta_e^2-1\right)=O(\delta)$. 
Therefore at leading order in $\delta$ we find, setting $F_* \approx 1$ as discussed in \cref{Error-Estimates}, that
\begin{equation}\begin{split}\label{main-result}
    P_\text{int}^{(0)}(E_+|q_0)&=
    \frac{1}{q_0 } \frac{\alpha}{4\pi}\int_{m_*^{(-)}}^{m_*^{(+)}} \frac{\dd m_*^2}{m_*^2} \times 2 \\
    &= \frac{1}{q_0}\frac{\alpha}{\pi} \log\qty(m_*^{(+)}\big/m_*^{(-)}) ~.
\end{split}\end{equation}
This function is plotted, along with error estimates discussed in \cref{Error-Estimates}, in \cref{main-plot}.  The small-$\delta$ form of $m_*^{(+)}/m_*^{(-)}$ depends on the relative size of $\epsilon$ and $\delta$.  Explicitly the various scaling limits are given by (always taking $\delta,\epsilon \ll 1$) 
\begin{equation}\label{ratio-cases}
    \frac{m_*^{(+)}}{m_*^{(-)}} 
    \sim 
    \begin{cases}
        2\delta /\epsilon & \delta \gg \epsilon \\ 
        \qty(\frac{\delta +\epsilon+\delta  \sqrt{1+2 \epsilon /\delta} }{\delta +\epsilon-\delta  \sqrt{1+2 \epsilon /\delta } })^{1/2} & \delta \sim  \epsilon\\
        1+ \sqrt{2 \delta/\epsilon} & \delta \ll \epsilon
    \end{cases}
\end{equation}
If one wishes to straddle these various regimes then the exact expressions for $m_*^{(\pm)}$, as presented in \cref{max-min-mstar}, should be used. This is important for applications at Mu2e and COMET where  energy resolution is of $O(100~\text{keV})$ \cite{Bernstein:2019fyh} such that electron mass effects can be resolved. 

\section{Conclusions and Outlook \label{Conclusions}} 

We have shown that given the photon spectrum from RMC on a nucleus, the spectrum of high energy positrons or electrons can be computed accurately with errors controlled by $\delta= (E_{-}-m_e)/q_0$. This allows for a robust characterization of the internal RMC positron and electron spectra near the end-point. External positrons, stemming from real photons pair producing in surrounding detector material, can be calculated from the real-photon spectrum by dedicated Monte Carlo simulations that include the full detector geometry. Thus, with this present work, a measurement of the real-photon spectrum becomes entirely predictive for the purposes of determining the high-energy positron spectrum. Most importantly, this allows for nuclear physics effects, which have large theoretical uncertainties, to be included empirically with measured data. 

While our study has emphasized the near end-point positron spectrum, it applies equally well to the near end-point electron spectrum.  The reason that electrons and positrons can be treated on an equal footing is that we have \emph{not} included the influence of the nucleus' strong Coulomb field, which, in reality, will influence the outgoing electron and positron. Near the end-point, for internal conversion, this approximate neglect of Coulomb corrections may be insufficient. A high-energy positron necessarily implies a low-energy electron whose outgoing wavefunction is then substantially distorted by the Coulomb field of the nucleus (as is well known in the theory of beta decay \cite{Wilkinson:1982hu}). This has been discussed in a more general setting in \cite{Schluter:1981cjo}, and we will study this issue in the context of the RMC end-point in future work \cite{Plestid-Future}. We anticipate that our formalism of an internal conversion probability can be readily adapted to account for Coulomb distortion effects. 

\section*{Acknowledgements} 

We are indebted to Robert Bernstein, Michael Mackenzie, Pavel Murat, and Stefano Di Falco for their consistent availability, enthusiasm,  and willingness to help us understand the details and needs of the Mu2e experiment. We especially thank Pavel Murat for emphasizing the importance of the internal positron spectrum and for the existing gap in the literature surrounding its near end-point behavior. This work was made possible by the Intensity Frontier Fellowship program which supported R.P.'s visit to Fermilab. R.P.\ is extremely grateful for the support and hospitality of the Fermilab theory group. This work was supported by the U.S. Department of Energy, Office of Science, Office of High Energy Physics, under Award Number DE-SC0019095. This manuscript has been authored by Fermi Research Alliance, LLC under Contract No. DE-AC02-07CH11359 with the U.S. Department of Energy, Office of Science, Office of High Energy Physics.

\appendix 

\onecolumngrid

\section{Error estimates \label{Error-Estimates}}

Here we consider corrections to the universal conversion probability $P_{\rm int}$. As emphasized in \cref{ratio-cases}, these corrections depend on the relative size of $\epsilon$ and $\delta$. In what follows we will consider the cases of $\epsilon \ll \delta$ [i.e.\ a relativistic electron satisfying $2 m_e \ll(E_\gamma - E_+)$], $\epsilon\sim \delta$ [i.e.\ a quasi relativistic electron satisfying $2 m_e \sim(E_\gamma - E_+)$] , and $\epsilon \gg \delta$ [i.e.\ a non-relativstic electron $2 m_e \gg(E_\gamma - E_+)$] separately.  

For this section it is convenient to re-express \cref{cos-theta-func} as 
\begin{equation}
    \cos\theta=\frac{2(1-\delta-\epsilon)-1}{\beta_*\beta_e}= \frac{1-2\delta-2\epsilon}{\beta_*\beta_e}~.
\end{equation}
where we remind the reader that $\beta_*=\sqrt{1-m_*^2/q_0^2}$, and $\beta_e= \sqrt{1-4m_e^2/m_*^2}$. The ratio of  $m_*^{(\pm)}$ to the photon energy $q_0$ will also be useful, and this can written in terms of $\beta$ and $\epsilon$ as
\begin{equation}\label{m-star-eps-delta}
    \frac{m_*^{(\pm)}}{q_0}=\sqrt{2\qty[\epsilon +\delta  (1-\delta -2 \epsilon )\pm \sqrt{\delta (1-\delta)  (1-\delta -2 \epsilon) (\delta +2 \epsilon )} ]} ~.
\end{equation}
Sub-leading corrections to $P_\text{int}$ depend on nuclear matrix elements and are therefore model dependent. We can, however, parameterize this model dependence at next to leading order by defining
\begin{align}        
    C_T(\Pi)&=q_0^2    
  \qty[\frac12\pdv[2]{m_*}~ T_*^2]_{m_*=0}~,\\       
    C_L(\Pi)&= \qty[L_*^2]_{m_*=0}~,
\end{align}             
where $C_T$ and $C_L$ depend on the other phase-space variables $\Pi$ ($\Pi=\cos\theta_{\nu\gamma}$, the opening angle between the photon and neutrino, for RMC). In estimating errors in the main text we treat $C_L$ and $C_T$ as constants independent of $\cos\theta_{\nu\gamma}$.  The small-$m_*$ behavior of the functions $T_*^2$ and $L_*^2$ can be expressed in terms of $C_T$ and $C_L$ via 
\begin{align}
    T_*^2 &= C_T(\Pi) \times \qty(\frac{m_*}{q_0})^2  + O(m_*^4)~,\\
    L_*^2 &= C_L(\Pi) + O(m_*^2) ~.
\end{align}
Depending on the relative sizes of $m_*^{(\pm)}$, either of $\beta_e$ or $\beta_*$ can deviate from unity at next to leading order. We therefore expand both of them in a Taylor series:
\begin{align}
    \beta_e&=\sqrt{1-\frac{4m_e^2}{m_*^2}}\sim 1-\frac{2m_e^2}{m_*^2}~,\\
    \beta_*&= \sqrt{1-\frac{m_*^2}{q_0^2}}\sim 1-\frac{m_*^2}{2q_0^2}~.
\end{align}
While the expansion in $m_e^2/m_*^2$ may look ill-behaved for small $m_*$, recall that $m_* \geq 2 m_e$, and for $\delta \ll 1$  we always have that $m_e/m_* \lesssim O(\sqrt{\delta})$. Throughout the full region of phase space we find that $F_* \sim 1 + O(\frac{m_*^2}{q_0 \MoneS})$ and for medium-heavy nuclei these deviations from unity are at the permille level since $m_*\sim 10 $ MeV,  $q_0 \sim 100$ MeV, and $\MoneS\sim 30$ GeV; we therefore set $F_*=1$ hereafter. 

We may then expand \cref{P-int-def} under the integral sign subtracting off the leading-order expression. We therefore define 
\begin{equation}
    \Delta P_\text{int} = P_\text{int}-P_\text{int}^{(0)}~,
\end{equation}
given explicitly by 
\begin{equation}\begin{split}
    \Delta P_\text{int} &= \frac{\alpha}{4\pi} \int \frac{\dd m_*^2}{m_*^2} \qty[ 2 C_T(\Pi) \frac{m_*^2}{q_0^2} -  \qty(4(\delta+\epsilon)-\frac{m_*^2}{q_0^2} -\frac{4 m_e^2}{m_*^2}) +2\qty(4(\delta+\epsilon)- \frac{m_*^2}{q_0^2}) C_L(\Pi) ]\\
    &=\frac{1}{q_0}\frac{\alpha}{\pi}\int \frac{\dd m_*}{m_*} \qty{  2[\delta + \epsilon][2C_L(\Pi)-1]+ [ C_T(\Pi)-C_L(\Pi)+\tfrac12] \frac{m_*^2}{q_0^2} +\frac{2m_e^2}{m_*^2}  }~. 
\end{split}\end{equation}
In deriving the above equation we have made use of 
\begin{align}
    (1-\cos^2\theta)\beta_e^2 &=\frac{ \beta_e^2\beta_*^2-(1 - 2\delta -2\epsilon)^2 }{\beta_*^2} \sim 4(\delta+\epsilon) -\frac{m_*^2}{q_0^2}-\frac{4m_e^2}{m_*^2}
    + O(\delta^2)~,\label{a6} \\
   (1- \cos^2\theta \beta_e^2) &=  
    \frac{\beta_*^2-(1 - 2\delta -2\epsilon)^2}{\beta_*^2}\sim 4(\delta +\epsilon) -\frac{m_*^2}{q_0^2}
    + O(\delta^2)~\label{a7}.
\end{align}
If we integrate over $m_*$ we then find 
\begin{equation}\begin{split}\label{above}
    \Delta P_\text{int} &= \frac{1}{q_0}\frac{\alpha}{\pi}\bigg\{  2[\delta + \epsilon][2C_L(\Pi)-1] \log(m_*^{(+)}/m_*^{(-)})+ [ \tfrac12C_T(\Pi)-\tfrac12 C_L(\Pi)+\tfrac14] \frac{[m_*^{(+)}]^2-[m_*^{(-)}]^2}{q_0^2} \\
    &\hspace{300pt}+\qty[\qty(\frac{m_e}{m_*^{(-)}})^2 - \qty(\frac{m_e}{m_*^{(+)}})^2]   \bigg\}~.
\end{split}\end{equation}
Different terms in \cref{above} will be relevant or negligible depending on whether $\epsilon \ll \delta$,  $\epsilon \sim \delta$, or $\epsilon \gg \delta$. It is therefore useful to consider these limits separately. 

\subsection{Relativistic electron: $\epsilon \ll \delta$} 
In this section we will power-count with $\epsilon \sim O(\delta^2)$ such that
\begin{align}
    \frac{m_*^{(+)}}{q_0} &\sim 2 \sqrt{\delta} ~,\\
   \frac{m_*^{(-)}}{q_0} &\sim \frac{\epsilon}{\sqrt{\delta}} \sim O(\delta^{3/2})~.
\end{align}
We then find 
\begin{equation}\label{case-1}
	\Delta P_\text{int} = \frac{2\delta}{q_0} \frac{\alpha}{\pi} \qty{[2C_L(\Pi)-1] \log(2\delta/\epsilon) + [C_T(\Pi) -C_L(\Pi)+1]   }~, 
\end{equation}
where all contributions proportional to $\epsilon$  have been dropped because of the $\epsilon \sim \delta^2$ power counting. The final term in \cref{case-1} receives a contribution from $[m_e/m_*^{(-)}]^2\sim O(\delta)$ and $[m_*^{(+)}/q_0]^2 \sim O(\delta)$.

\subsection{Quasi-relativistic electron: $\epsilon \sim \delta$} 

In this section we treat $\epsilon \sim O(\delta)$ in our power counting, and find 
\begin{align}
   \frac{m_*^{(\pm)}}{q_0} &\sim  \sqrt{2\delta}\times\qty[1+\frac{\epsilon}{\delta} \pm \sqrt{1+\frac{2\epsilon}{\delta}}]^{1/2}~.
\end{align}
Notice that both $m_*^{(+)}$ and $m_*^{(-)}$ are parametrically of the same size, in contrast to the case where $\epsilon \ll \delta$ where we instead found that $m_*^{(+)} \gg m_*^{(-)}$. We then find 
\begin{equation}\label{case-2}
		\Delta P_\text{int} =  \frac{2\delta}{q_0} \frac{\alpha}{\pi}  \qty{[2C_L(\Pi)-1](1+\tfrac{\epsilon}{\delta})\log(m_*^{(+)}/m_*^{(-)}) + \sqrt{1+\tfrac{2\epsilon}{\delta}}[C_T(\Pi) -C_L(\Pi)+1]   }~, 
\end{equation}
where, for compactness, we have left the dependence of $m_*^{(\pm)}$  on $\epsilon$ and $\delta $ implicit inside the logarithm. The final term in \cref{case-2} receives contributions from both roots of $m_*^{(\pm)}$. 

\subsection{Non-relativistic electron: $\epsilon \gg \delta$}

In this section $\epsilon \sim O(\delta^{1/2})$ in our power counting, and
\begin{align}
    \frac{m_*^{(\pm)}}{q_0} &\sim \sqrt{2\epsilon}\pm \sqrt{\delta} ~.
\end{align}
Much like in the previous section both roots are of roughly the same size. We then find that 
\begin{equation}\begin{split}\label{case-2}
	\Delta P_\text{int} &= \frac{\sqrt{8\delta\epsilon}}{q_0} \frac{\alpha}{\pi}  [C_T(\Pi)+C_L(\Pi)]~,
\end{split}\end{equation}
where we have Taylor expanded $\log(m_*^{(+)}/m_*^{(-)})$ taking advantage of the fact that $\delta \ll \epsilon$. In this regime the error scales parametrically like $O(\sqrt{\epsilon\delta})$. This should be compared to $P_\text{int}^{(0)}$ in the same regime which is $O(\sqrt{\delta/\epsilon})$. The relative error is therefore $O(\epsilon)$. This stems from the photon virtuality $m_*$ tending towards $m_*^2 \rightarrow 2 m_e q_0$ in the limit that $\delta \rightarrow 0$ such that $m_*^2/q_0^2 \sim O(m_e/q_0) \sim O(\epsilon)$. 

\section{Virtuality dependence of the transverse electromagnetic current \label{OFPT} }

In \cref{p-int-approx} we have assumed that corrections stemming from $L_*^2$ and $T_*^2$ are $O(\delta)$ or smaller.  The longitudinal matrix elements, $L_*^2$, are suppressed explicitly by a factor of $(1-\beta_e^2\cos^2\theta)\sim O(\delta)$. In contrast, the transverse matrix elements, $T_*^2$, are not suppressed by explicitly $O(\delta)$ prefactors. For corrections from $T_*^2$ to be small, we therefore require that terms proportional to the virtual photon mass $m_*^2$ are suppressed by energetic scales of $O(q_0^2)\sim O(m_\mu^2)$ such that $T_*^2\sim m_*^2/q_0^2\sim O(\delta)$.  One may be concerned that energy-level splittings (on the order of $\Delta E\sim$ few MeV) from low-lying nuclear excited states  could enhance $m_*$ dependent contributions leading to $O(m_*^2/\Delta E^2)$ corrections. In this section we study the structure of $\mathscr{J}_*^{\mu\nu}$ by inserting a complete set of nuclear states, and deriving a time-ordered expression where energy-splitting denominators appear explicitly.

We will study the matrix element 
\begin{equation}
	_{\text{out}}\hspace{-4pt}\bra{\nu, A'} \Jem \ket{ \AmuAtom}_\text{in}\approx \int \frac{\dd^3 k}{(2\pi)^3} \tilde{\psi}_\mu(k) _{\text{out}}\hspace{-4pt}\bra{\nu, A'} \Jem \ket{ A(-k),\mu(k)}_\text{in}~ \approx \psi_\mu(0) _{\text{out}}\hspace{-4pt}\bra{\nu, A'} \Jem \ket{ A,\mu}_\text{in}~ 
\end{equation}
where we have assumed a non-relativistic treatment of the muon-nucleus bound state in the first approximation, and  approximated the plane-wave matrix element by its value at $k=0$ in the second approximation; we have also used $\psi_\mu(0)=\int \tilde{\psi}_\mu(k) \dd^3 k/(2\pi)^3$.

We are interested in this matrix element for real- and virtual-photon kinematics at fixed photon energy $q_0$. We take the initial state as having zero three-momentum and an energy (i.e.\ rest mass) of $\MoneS$. The outgoing momentum of $A'$ is fixed by momentum conservation, and the energy of the neutrino by energy conservation. In the limit where $M_{A'} \sim\MoneS \rightarrow \infty$ (i.e.\ neglecting the recoil energy of $A'$) we find
\begin{equation}\begin{split}\label{neutrino-energy}
	E_\nu &= (\MoneS-M_{A'} ) -q_0  + O(q_0^2/M_{A'})\\
    			&\approx m_\mu - q_0 ~,
\end{split}\end{equation}
where in the approximation we have neglected the binding energy of the muon and the mass difference $M_{A'}-\MoneS$. 

Let us evaluate $_\text{out}\hspace{-2pt}\langle \hat{\mathscr{J}}_\mu \rangle_\text{in}$ at leading order in $G_F$ using Dyson's formula, such that the lepton and nuclear matrix elements can be fully separated
\begin{equation}\begin{split}
  _{\text{out}}\hspace{-4pt}\bra{\nu, A'} \Jem \ket{ A,\mu}_\text{in}
  &= \bra{\nu, A'}T \{ \e^{-\iu \int \dd^4 y \mathcal{H}_\text{int}(y) } \Jem(0) \}\ket{ \mu, A}\\
  &=\frac{G_F}{\sqrt{2}}  \iu\int \dd^4 y   \bra{\nu, A'}T \{ \bar{\nu}(y)\gamma^\nu(1-\gamma_5)\ell(y) \Jcc(y) \Jem(0) \}\ket{ \mu, A}  ~ + O(G_F^2)~,
\end{split}\end{equation}
where $\hat{J}_\nu$ is the weak-hadronic current. Furthermore, the electromagnetic current is a linear combination of a leptonic and hadronic piece
\begin{equation}\label{linearity-current}
  \hat{\mathscr{J} }_\mu = \hat{\mathscr{H}}_\mu + \sum_\ell -e \bar{\ell} \gamma_\mu \ell~. 
\end{equation}
We can therefore write (dropping in- and out-labels since we are now calculating perturbatively)
\begin{equation}\begin{split}\label{dyson}
    \bra{\nu, A'} \Jem \ket{ \mu, A}
      = \frac{G_F}{\sqrt{2}}  \int \dd^4 y
      \bigg[-e\bra{\nu} T\{\bar{\nu}(y)&\gamma^\nu (1-\gamma_5)\ell(y)
        \bar{\ell}(0)\gamma^\nu\ell(0) \} \ket{\mu}
        \bra{A'} \Jcc(y) \ket{A} \\
        &+ \bra{\nu}\bar{\nu}(y)\gamma^\nu (1-\gamma_5)\ell(y)\ket{\mu}
        \bra{A'}T\{ \Jcc(y)   \Jem(0) \}\ket{A}\bigg]~,
      \end{split}
\end{equation}
where we have used \cref{linearity-current} and the fact that $\bra{A'}T\{ \Jcc(y)   \Jem(0) \}\ket{A}= \bra{A'}T\{ \Jcc(y)   \hat{\mathcal{H}}_\mu(0) \}\ket{A}$.

The first term can be reduced to a single weak-current nuclear matrix element (measurable with e.g.\ neutrino scattering)  and an electroweak Feynman diagram with the muon radiating a photon. In the second term the nucleus radiates a photon and is excited by a weak current insertion. In between these two current insertions a virtual excitation of some low-lying nuclear energy level could, naively, supply an energetic denominator that is $\Delta E \sim O(\text{few MeV})$. Our focus therefore shifts to this term. 

Translating the lepton fields to $t=0$ and  $y=0$ and introducing $Q_\ell =k-k'$ (the four momentum transferred out of the leptons) we see that all of the nuclear physics that could enhance terms proportional to $m_*^2$ is buried in the current-current correlation function evaluated between two nuclear states
\begin{equation}
  W_{\mu\nu} = \int \dd^4 y ~\bra{A'}
  T\{ \Jcc(y)\Jem(0)\}\ket{A} \e^{-\iu Q_\ell y} ~. 
\end{equation}
In terms of this object the second term in \cref{dyson} can be expressed as $ \frac{1}{\sqrt2}G_F \bar{u} \gamma^\nu (1-\gamma_5) u W_{\mu\nu}$.
Let us consider each time ordering separately:
\begin{align}
  \mathcal{W}_{\mu\nu}^+&= \int_0^\infty \dd y_0 \int \dd^3 y  ~\bra{A'}
  \Jcc(y)\hat{\mathscr{J}_\mu}(0)\ket{A} \e^{-\iu Q_\ell y}~, \\
    \mathcal{W}_{\mu\nu}^-&= \int_{-\infty}^0 \dd y_0 \int \dd^3 y  ~\bra{A'}
  \Jem(0) \Jcc(y)\ket{A} \e^{-\iu Q_\ell y}~.
\end{align}
Our approach is to:
\begin{enumerate}
\item Insert a complete set of energy eigenstates $\sumint \ketbra{X}=1$; we will refer to these states as isobars. 
\item Integrate over space to get a momentum conserving delta function.
\item Integrate over time to get denominators of $1/(E\pm \iu \eta)$. 
\end{enumerate}
We find 
\begin{align}
    \mathcal{W}_{\mu\nu}^+&=\sumint_X (2\pi)^3 \delta^{(3)}(\vec{Q}_\ell+ \vec{P}_X-\vec{p}\hspace{2.5pt}')  \frac{\bra{A'} \Jcc \ket{X} \bra{X} \Jem \ket{A}}{(Q_\ell+P_X -p')_0-\iu \eta}~,\\
    \mathcal{W}_{\mu\nu}^-&=-\sumint_{X} (2\pi)^3\delta^{(3)}(\vec{Q}_\ell-\vec{P}_X + \vec{p})  \frac{\bra{A'}\Jem\ket{X}\bra{X}\Jcc\ket{A} }{(Q_\ell+p-P_X)_0 +\iu \eta}~.
\end{align}
Next, we can make use of Lorentz invariance to re-write $\sumint_X =  \sum_n \int \widetilde{\dd p}_n $ where states are now labeled by their quantum number $n$ (which determines the mass and spin of the isobar state) and their three momentum $\ket{X}=|n,\vec{P}\rangle$; here $\widetilde{\dd p}= \tfrac{1}{2 E_p} \dd^3 p/(2\pi)^3$. Let us introduce $\ket{Z}= |n, \vec{P}=\vec{p}\hspace{1.5pt}'-\vec{Q}_\ell\rangle$ and $\ket{Y}=|n, \vec{P}=\vec{Q}_\ell-\vec{p}\rangle$ such that
\begin{equation}\label{wmunu-OFPT-xy}
  W_{\mu\nu} = \sum_Z  \frac{\bra{A'} \Jcc \ketbra{Z}{Z} \Jem \ket{A}}{2 E_Z (\Delta E_\ell+[E_Z-E_{A'}])-\iu \eta}~
-\sum_Y \frac{\bra{A'}\Jem\ketbra{Y}{Y}\Jcc\ket{A} }{2 E_Y (\Delta E_\ell-[E_Y-E_A]) +\iu \eta}~.
\end{equation}
We have introduced $\Delta E_\ell = (Q_\ell)_0$ and the energies of the isobars $E_Y$ and $E_Z$. Notice that the isobars $\ket{Z}$ and $\ket{Y}$ are discrete states whose momentum is fixed by three-momentum conservation, $\vec{P}_Z= \vec{p}\hspace{2.5pt}'-\vec{Q}_\ell$  and $\vec{P}_Y= \vec{Q}_\ell-\vec{p}$.  
Small nuclear energy level splittings from the ground state do not lead to small denominators in \cref{wmunu-OFPT-xy}. $\Delta E_\ell\sim O(q_0) \sim 100$ MeV is sufficiently large that any low-lying nuclear states are far off-shell.  The only small denominators that can appear are in the second time ordering when $[E_Y-E_A] \approx \Delta E_\ell$; such considerations are beyond the scope of this work, and we do not consider them here.

\section{Comparison with Kroll and Wada \label{KW-vs-HP}} 
The main difference between \cite{Kroll:1955zu} (KW) and our study is that we focus on the end-point \emph{specifically} and carefully consider the validity of approximating virtual photons by real ones. Since the authors of KW simply assume this approximation as an ansatz many of our results are similar to theirs. We disagree with the authors on certain technical points, and in particular on the validity of their advocated approximation in generic regions of phase space. We find, however, that the ansatz proposed in KW becomes a well controlled approximation in the limit that the electron's kinetic energy becomes much smaller than the virtual photon energy, i.e.\ when $\delta\ll 1$. 

The notation of KW differs substantially from modern treatments and specifically this paper. Moreover, they consider a two-body system and so details of the phase space are slightly different between our works. For the benefit of the interested reader we spell out the differences in notation, derive the results of KW, and comment on typos and conceptual differences. 

Kroll and Wada consider a system $A$, initially at rest, that emits a photon becoming system $B$ of mass $M$. The mass of system $A$ is parameterized by $M_A=M+E_{AB}$ with $E_{AB}$ the energy splitting between $A$ and $B$.  An important identity is that 
\begin{equation}
   \frac{1}{2 E_{B*}} =  \frac{E_{AB}+M}{(E_{AB}+M)^2+M^2-m_*^2}~.
\end{equation}
The case of a massless photon, $E_B$, is recovered by setting $m_*=0$.

In Eq.\ (2) of KW, the rate of photon emission is calculated. To emphasize the difference in definition we denote our equivalent expression by (P\&H) and the expressions of KW by (K\&W),
\begin{align}
    \Gamma_\gamma&= \frac{1}{8\pi^2} \frac{q_0 M_A }{ E_B} \int \dd \Omega \qty(J_0^{11}+J_0^{22}) \qq{(K\&W)}~,\\
    \Gamma_\gamma&= \frac{1}{2M_A} \times \frac{\bar{\beta}(M_A, 0,M_B)}{32\pi^2} \int \dd \Omega \qty(\mathscr{J}_0^{11}+\mathscr{J}_0^{22}) \qq{(P\&H)}~,
\end{align}  
where (borrowing notation from \cite{HitoshiPS}) 
\begin{equation}
    \bar\beta(s,m_1,m_2)= \sqrt{1-\frac{2(m_1^2+m_2^2)}{s} + \frac{(m_1^2-m_2^2)^2}{s^2}}~.
\end{equation}
For $m_1=0$ we have that $\bar\beta=2 E_1/M_A$ such that $\bar\beta(M_A,0,M_B)= 2 q_0/M_A$ and therefore 
\begin{equation}
     \Gamma_\gamma=\frac{1}{2 M_A} \frac{1}{16\pi^2} \frac{q_0}{ M_A } \int \dd \Omega \qty(\mathscr{J}_0^{11}+\mathscr{J}_0^{22})\qq{(P\&H)}~.
\end{equation}
This allows us to identify the (K\&W) current with our (P\&H) matrix element via  
\begin{equation}\label{J-vs-J}
    \qq{(K\&W)} J_\mu = \frac{\sqrt{E_B}}{2 M_A} \langle\mathscr{J}_\mu\rangle \qq{(P\&H)}~.
\end{equation}
This means that there are implicit factors of $E_B$ contained in the expressions in KW, with $J_*^{\mu\nu}\propto E_{B*}\times \mathscr{J}_*^{\mu\nu}$. 

Having identified the appropriate current we can now jump ahead to the internal conversion coefficient $\rho$ which is defined as the rate of electron-positron pair production to photon emission. KW gives (with expressions adapted to our notation) in Eq.\ (8) 
\begin{equation}
    \rho= \frac{\alpha}{2\pi} \int_{2m_e}^{q_0} \frac{\dd m_*}{m_*} \int_{-1}^1\dd\cos\vartheta~\frac{E_{B*}}{E_B}  \frac{|\vec{q}_*|}{|\vec{q}|}   \beta_e \qty{ \qty[2+2(\cos^2\vartheta-1)\beta_e^2]R_T(m_*) + 2\qty[1-\beta_e^2\cos^2\vartheta] \frac{m_*^2}{q_0^2} R_L(m_*)}~\qq{(K\&W)},
\end{equation}
where we have used 
\begin{equation}
    q_0= \frac{E_{AB}^2+2 E_{AB}M+m_*^2}{2 (E_{AB}+M)}~.
\end{equation}
Modifying  our discussion in the main text to account for two-body phase space we find that 
\begin{equation}
    \Phi_{2*} = \Phi_2 \times \frac{\bar\beta_*}{\bar\beta_0} ~,
\end{equation}
with 
\begin{equation}
    \bar\beta_*= \sqrt{1-\frac{2(M_B^2+m_*^2)}{M_{A}^2} + \frac{(M_B^2-m_*^2)^2}{M_A^4}}~.
\end{equation}
One can check that $|\vec{q}_*|/|\vec{q}_0|= \bar\beta_*/\bar\beta_0$. We therefore find [using $\dd m_*^2 = 2 m_* \dd m_*$ and keeping in mind that we are \emph{not} including the delta function $\delta(\mathfrak{E}_+-E_+)$], 
\begin{equation}
     \rho= \frac{\alpha}{2\pi} \int_{2m_e}^{q_0} \frac{\dd m_*}{m_*} \int_{-1}^1\dd\cos\vartheta~~~ \frac{|\vec{q}_*|}{|\vec{q}|} \times  \beta_e \qty{ \qty[2+2(\cos^2\vartheta-1)\beta_e^2](1+T_*^2) + 2\qty[1-\beta_e^2\cos^2\vartheta] L_*^2}~\qq{(P\&H)}.
\end{equation}
The factor of $E_{B*}/E_{B}$ obtained in KW is absent from our expression because our functions $T_*^2$ and $L_*^2$ are defined in terms of $\mathscr{J}$ whereas $R_T$ and $R_L$ are defined in terms of KW's $J$. As emphasized in \cref{J-vs-J}, upon being squared these differ by a factor of $E_B$, and hence in KW's expression there \emph{should} be a ratio of $E_{B}/E_{B*}$ instead of the factor of $E_{B*}/E_{B}$ that appears in Eqs.\ (8) and (9) of KW (explicitly, $[(E+M)^2+M^2-x^2]/[(E+M)^2+M^2]$ in the notation of KW); we ascribe this to a typo in KW.

Despite appearances there is no discrepancy between our longitudinal matrix element and those in KW, because KW have evaluated their matrix elements in the lab frame, whereas our expression is evaluated in the rest-frame of the virtual photon (or equivalently the electron positron pair). This accounts for the factor of $m_*^2/q_0^2$ in KW, which can be interpreted as arising from boosting the rest-frame expression into the lab-frame. 

This final point is important since the naive $m_*^2$ suppression appearing in front of $R_L(m_*)$ is artificial, but was used as a justification for the neglect of longitudinal matrix elements in KW. The fact that this is an artificial suppression can be seen in a simple example such as $\ell^+\ell^- \rightarrow e^+ e^-$ where we can interpret the $\ell^+\ell^-$ pair as a current sourcing a virtual photon $\gamma_*$. In this case a straightforward calculation shows that $\mathscr{J}_*^{\mu\nu}=(k^\mu p^\nu +p^\mu k^\nu - \tfrac12 s g_{\mu\nu})$ such that $\mathscr{J}_*^{33}=2 m_\ell^2$ in the rest-frame (i.e. center of mass frame), whereas it equals $\mathscr{J}_*^{33}=2\gamma_*^2 m_\ell^2$ in the lab frame, where $\gamma_* = q_0^2/m_*^2$ is the boost of the virtual photon. Note, however, that in \emph{any frame} the combination $m_*^2/q_0^2 \times \mathscr{J}_*^{33}=2 m_\ell^2 $, which does \emph{not} vanish in the $m_*\rightarrow 0$ limit. This illustrates that longitudinal matrix elements do not decouple in the $m_*\rightarrow 0$ limit. Instead, as we have shown in \cref{lepton-tensor,exact-rate-1,P-int-def}, they decouple when $\beta_e\cos\theta\rightarrow 1$. 

The neglect of finite-$m_*^2$ corrections to the transverse matrix elements is only valid if $m_*^2 \ll q_0^2$. The integration measure is logarithmic in $m_*$, and so all scales contribute equally. Therefore, unless $(m_*^{(+)}/q_0)^2 \ll  1$, there is no reason to expect a real-photon approximation to be accurate. Additionally, as we have emphasized in the paragraph above, there is no small-virtuality suppression of longitudinal matrix elements. Rather, these are suppressed when $[1-\beta_e^2\cos^2\vartheta]$ is small.  As the positron energy approaches its endpoint, both of these conditions are satisfied with errors being $O(\delta)$.

\section{Generalization to arbitrary final states \label{inclusive-X}} 

In \cref{reaction} we have assumed a definite exclusive final state. As mentioned in the first footnote of the paper this assumption can be easily relaxed to accommodate inclusive final states for reactions of the form
\begin{equation}
    \ket{\AmuAtom}\rightarrow \sum_X~\ket{\nu,\gamma, X}~.
\end{equation}
The expressions for the real photon spectrum and internal $e^+e^-$ spectrum are given by 
\begin{align}
      \Gamma_{\gamma}&=\frac{1}{2\MoneS} \sum_X \int \dd \Phi_{3}(X)  \mathscr{J}_0^{\mu\nu}(X)(-g_{\mu\nu}^\perp)~,\\
    \Gamma_{e_+e_-}&=\frac{1}{2\MoneS} \sum_X \int \dd \Phi_{4}(X) \mathscr{J}_*^{\mu\nu}(X) \frac{4\pi\alpha}{m_*^4}L_{\mu\nu} ~,
\end{align}
where we have included an $X$ dependent evaluation of the electromagnetic current.  Four-body phase space can be decomposed for each final state $X$ as described in the main text.  If $X$ is has a multi-particle final state then an analogous decomposition for $n$-body phase space can be performed. 

Working at leading order in $\delta$ the resultant $e^+e^-$ spectrum can then be related to the photon spectrum by the same internal conversion probability for every final state $X$
\begin{equation}
    \dv{\Gamma_{e_+e_-}}{E_+}=  \int \dd E_\gamma \sum_X\dv{\Gamma_X}{E_\gamma} P_\text{int}^{(0)}(E_+|E_\gamma)~. 
\end{equation}
The full differential photon flux is then easily identified  as $\dd \Gamma= \sum_X \dd \Gamma_X$, such that
\begin{equation}
    \dv{\Gamma_{e_+e_-}}{E_+}= \int \dd E_\gamma \dv{\Gamma}{E_\gamma} P_\text{int}^{(0)}(E_+|E_\gamma)~. 
\end{equation}
This agrees with \cref{P-int-dE}. 

\vfill 
\pagebreak

\bibliography{bibfile.bib}
\end{document}